\documentclass[10pt]{article}
\usepackage{graphicx}
\usepackage{hyperref}
\usepackage{wrapfig}
\usepackage{lineno}

\def\Title#1{\begin{center} {\Large #1 } \end{center}}
\def\Author#1{\begin{center}{ \sc #1} \end{center}}
\def\Address#1{\begin{center}{ \it #1} \end{center}}

\newcommand\pubblock{\rightline{\begin{tabular}{l} Proceedings of the Fifth Annual LHCP\\ \pubnumber\\
         \pubdate  \end{tabular}}}

\newenvironment{Abstract}{\begin{quotation} \begin{center} 
             \large ABSTRACT \end{center}\bigskip 
      \begin{center}\begin{large}}{\end{large}\end{center} \end{quotation}}

\newenvironment{Presented}{\begin{quotation} \begin{center} 
             PRESENTED AT\end{center}\bigskip 
      \begin{center}\begin{large}}{\end{large}\end{center} \end{quotation}}

\def\beq{\begin{equation}}
\def\eeq#1{\label{#1}\end{equation}}
\def\eeqn{\end{equation}}

\def\beqa{\begin{eqnarray}}
\def\eeqa#1{\label{#1}\end{eqnarray}}
\def\eeqan{\end{eqnarray}}

\let\bar=\overbar

\def\Dslash{\not{\hbox{\kern-4pt $D$}}}
\def\dslash{\not{\hbox{\kern-2pt $\del$}}}

\def\msb{{\bar{\ssstyle M \kern -1pt S}}}

\usepackage{color}

\newcommand\pubnumber{ ATL-PHYS-PROC-2017-146 }
\newcommand\pubdate{\today}

\def\affiliation{
On behalf of the ATLAS and CMS Collaborations, \\
Department of Physics \\
Stony Brook University, Stony Brook, USA}

\begin{document}
%\linenumbers
% large size for the first page
\large
\begin{titlepage}
\pubblock

%% Change the title, name, abstract
%% Title 
\vfill
\Title{  Searches for new resonances decaying to W, Z and H bosons
  with the   ATLAS and CMS detectors in 13 TeV proton-proton
  collisions at the LHC  }
\vfill

%  if you need to add the support use this, fill the \support definition above. 
%   \Author{ FIRSTNAME LASTNAME \support }
\Author{ LJILJANA MORVAJ  }
\Address{\affiliation}
\vfill
\begin{Abstract}
Searches for high-mass resonances decaying to pairs of standard model bosons using ATLAS and CMS detectors are
presented. The searches use 15 - 36 $\textrm{fb}^{-1}$ of proton-proton collision data collected
at the centre-of-mass energy of 13 TeV at the LHC. Many different final states have been considered and the results are
interpreted in a range of theories beyond the standard model of particle physics, most common being heavy
vector triplet and warped extra dimensions.
\end{Abstract}
\vfill

% DO NOT CHANGE 
\begin{Presented}
The Fifth Annual Conference\\
 on Large Hadron Collider Physics \\
Shanghai Jiao Tong University, Shanghai, China\\ 
May 15-20, 2017
\end{Presented}
\vfill
\end{titlepage}
\def\thefootnote{\fnsymbol{footnote}}
\setcounter{footnote}{0}
%

% normal size for the rest
\normalsize 

%% Your paper should be entered below. 

\section{Introduction}

Many theories of physics beyond the standard model (BSM) predict
the existence of new heavy particles that decay to pairs of standard
model (SM) bosons, either vector bosons (V), being W and Z, or the Higgs boson (H).
BSM models most commonly used for the interpretation of the results are the heavy
vector triplet (HVT) \cite{HVT} and bulk Randall-Sundrum (RS) scenario of
warped extra dimensions \cite{RS}. 
HVT is  a simplified model  used to describe the phenomenology of new spin-1 resonances (W'
and Z') with a
small number of parameters. An interesting case for this report is the
so-called scenario B where the fermionic couplings are suppressed and
decays to bosons dominate. In the bulk scenario of the Randall-Sundrum
model, all the SM fields, along with the
spin-2 graviton, are allowed to propagate in the warped extra
dimension. In this RS scenario the couplings of the graviton to light fermions are
suppressed (and the coupling to photons negligible) making decays to
dibosons a highly motivated case.
The searches for high mass VH, VV and HH resonances have been performed in a large number of final states,
however, this report focuses only on a couple of recent analyses with
15-36 $ \textrm{fb}^{-1}$ of the 13 TeV ATLAS \cite{ATL} and CMS
\cite{CMS} datasets.

\section{Common analysis techniques}

Standard model bosons emerging from the decays of a heavy BSM state
(where heavy means having a mass much larger than the mass
scale of SM states) are often
highly boosted. If the momentum of the  SM boson is sufficiently high, the angular separation between
its hadronic decay products will be smaller than angular size
typically used by the experiments for the jet
reconstruction, resulting in the boson being reconstructed as a single
large radius (fat) jet.
ATLAS and CMS have developed sophisticated techniques for reconstructing
and tagging such boosted states. Jet grooming is used to remove soft
and large angle QCD radiation and pile-up contributions, and to improve the resolution
of V/H-jet mass. Jet substructure and tagging techniques are then used
to further discriminate between the background quark/gluon-initiated jets and signal V/H-jets.

\subsection{Jet grooming}
ATLAS reconstructs fat jets using
topological clusters as the input to the anti-kT algorithm with the
angular parameter R=1.0. A trimming technique \cite{Trim} is used to
remove the pileup contamination --  jet constituents are
reclustered into subjets with R=0.2 and all the subjets
satisfying the condition
$p_{\textrm{T}}(\textrm{subjet})/p_{\textrm{T}}(\textrm{jet})<0.05$
are removed.  Jet mass is then computed
using information from both the calorimeter and the tracks associated
with the jet \cite{CombMass}.

CMS reconstructs particle-flow jets using the anti-kT algorithm with the angular parameter R=0.8. Pileup per particle identification \cite{PUPPI} algorithm uses a
weight assigned to each particle based on its likelihood to originate from
pileup interactions to rescale the particle 4-momenta. 
A modified mass-drop algorithm, referred to as soft-drop algorithm
\cite{SoftDrop1, SoftDrop2, SoftDrop3},  iteratively breaks the jet into 2 sub-jets and drops the softer one until
the soft-drop condition is satisfied \cite{CMSdrop}.
This procedure improves the resolution in the V jet mass, lowers the mass of
quark/gluon-initiated jets from multijet background and further
reduces the pileup contamination.

\subsection{Boosted boson tagging}

ATLAS tags Higgs boson decays to two b-quarks by requiring one or two
 b-tagged track jets (R=0.2)  associated to a fat jet consistent with
 the Higgs boson mass. Boosted W/Z-boson candidates are required to
 have their mass consistent with the W/Z-mass within window size ranging
 between 15 GeV and 40 GeV, depending on the analysis. The ratio of energy
 correlation functions $D_2^{\beta=1}$  \cite{D2_1, D2_2} is used to
 test the compatibility with a two-prong decay topology of the fat
 jet. Transverse momentum dependent
 selection on the $D_2^{\beta=1}$ variable is applied with the approximate
 V-tagging efficiency of 50\% and quark/gluon-jet mistag efficiency of
 2\%.

CMS uses a dedicated b-tagging discriminator to identify two b-quarks
clustered in a single jet with mass consistent to that of the Higgs boson.
Boosted V-candidates are required to have their mass consistent either
with W (65 GeV -- 85 GeV) or Z (85 GeV -- 105 GeV) mass. The ratio of N-subjettiness \cite{tauN}
variables, $\tau_{21}=\tau_2/\tau_1$, that quantify the capability of reclustering the jet constituents in
exactly two ($\tau_2$) or one ($\tau_1$) subjets is used to assess the
compatibility with a two-prong decay.

\section{VH $\rightarrow$ qqbb}
\label{sec:VH}
\begin{figure}[htb]
\centering
\includegraphics[height=2.2in]{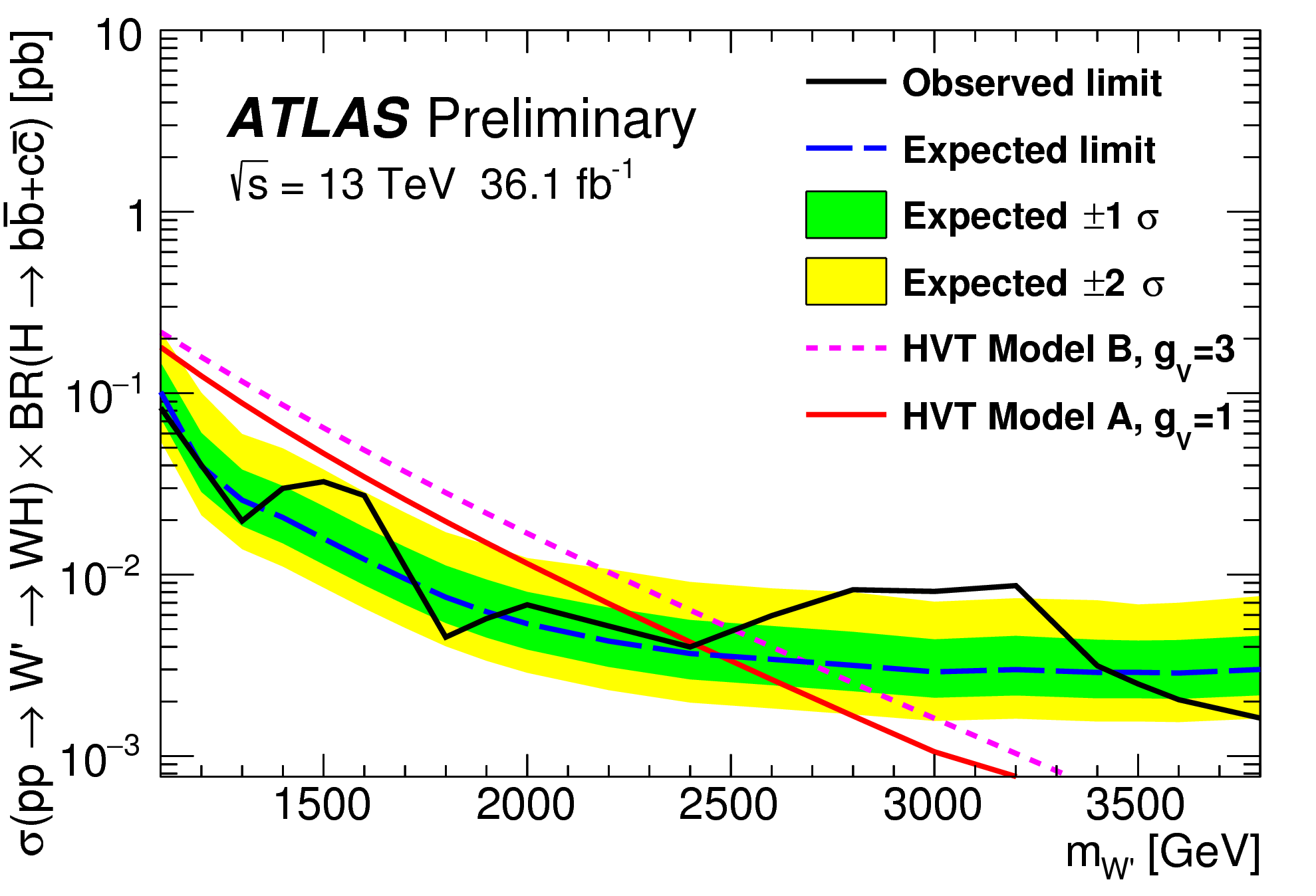}
\includegraphics[height=2.3in]{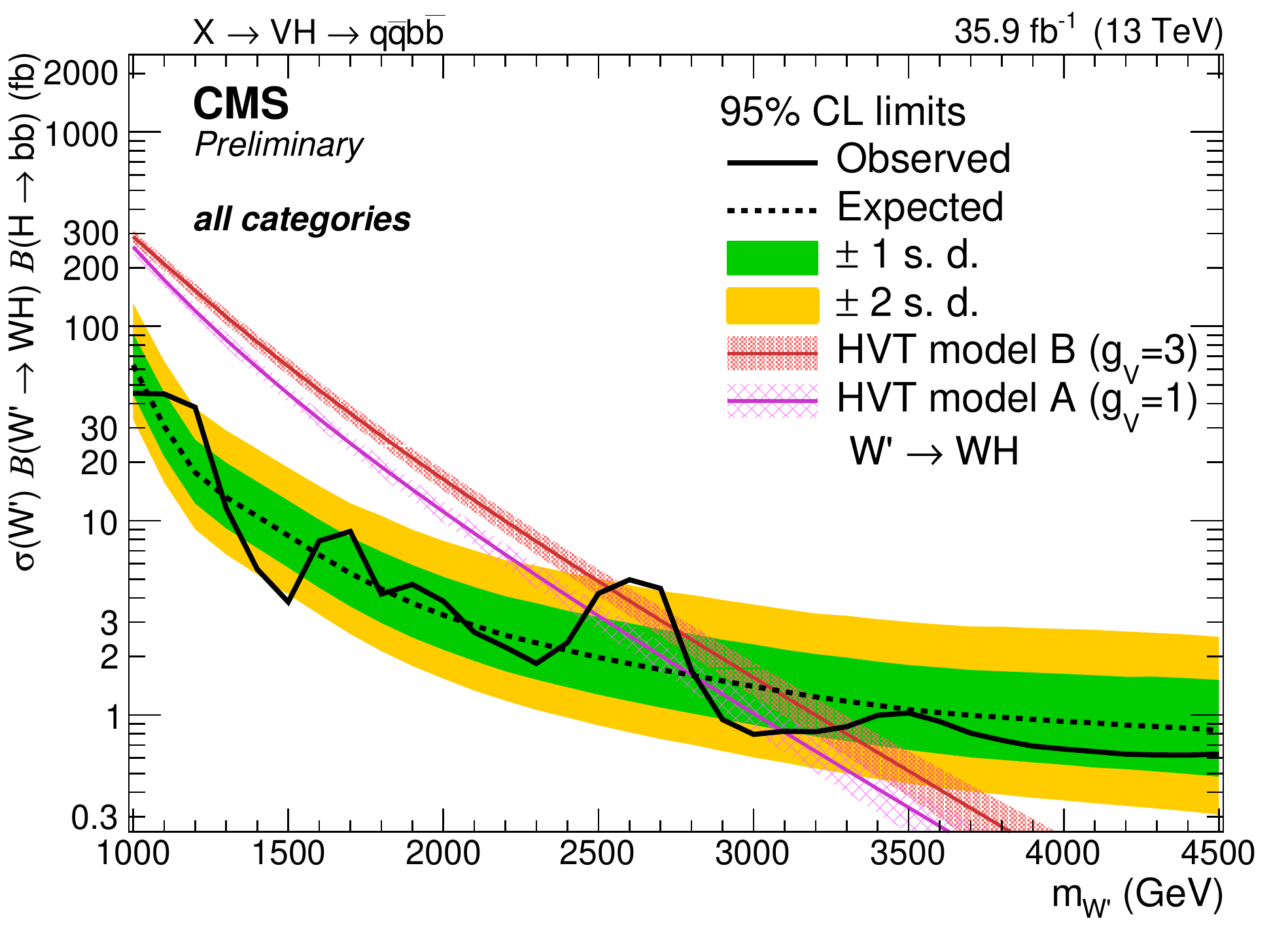}
\caption{Observed (solid black) and expected (dashed) limits at
  95\% CL on the cross section times branching ratio for the W'$\rightarrow $ WH search in ATLAS (left)
  \cite{ATLVH} and CMS
  (right) \cite{CMSVH}. The red and magenta curves show the predicted
  cross-sections as a function of resonance mass for the A and B
  scenarios of the HVT model.}
\label{fig:VH}
\end{figure}

ATLAS analysis \cite{ATLVH}  searches for a resonance in the invariant mass spectrum
of two fat jets, $m_{JJ}$.
Four search regions are defined based on the number of
b-tagged track jets associated with H-candidate and the mass of the
V-candidate. The dominant multijet background is estimated using a
template extracted from the data in the region with zero b-tagged
track jets associated to the H-candidate. The template is corrected with kinematic
reweighting in order to correctly reproduce the kinematics of the
search region and validated in the V-mass sidebands. Upper limits are set on the production cross section times the
branching ratio to VH(H$\rightarrow$heavy-flavour jets) final statesr
in the  $1<m_{JJ}<3.8$ TeV region (Fig. \ref{fig:VH}, left). The
biggest discrepancy between the SM expectation and the observed data
is found at $m_{JJ}\sim 3$ TeV and has the global significance of 2.2
$\sigma$.

The corresponding CMS analysis \cite{CMSVH} considers eight exclusive search regions
depending on the V-jet mass, H-jet b-tagging discriminator and
$\tau_{21}$ value for V-jets. The dominant multijet background is extracted in a fit to the data with an analytic
function and validated in the V-mass sidebands. Limits are set in $1<m_{JJ}<4.5$ TeV region
(Fig. \ref{fig:VH}, right). No significant excess is found.

\section{VV $\rightarrow$ qqqq}
\begin{figure}[htb]
\centering
\includegraphics[height=2.2in]{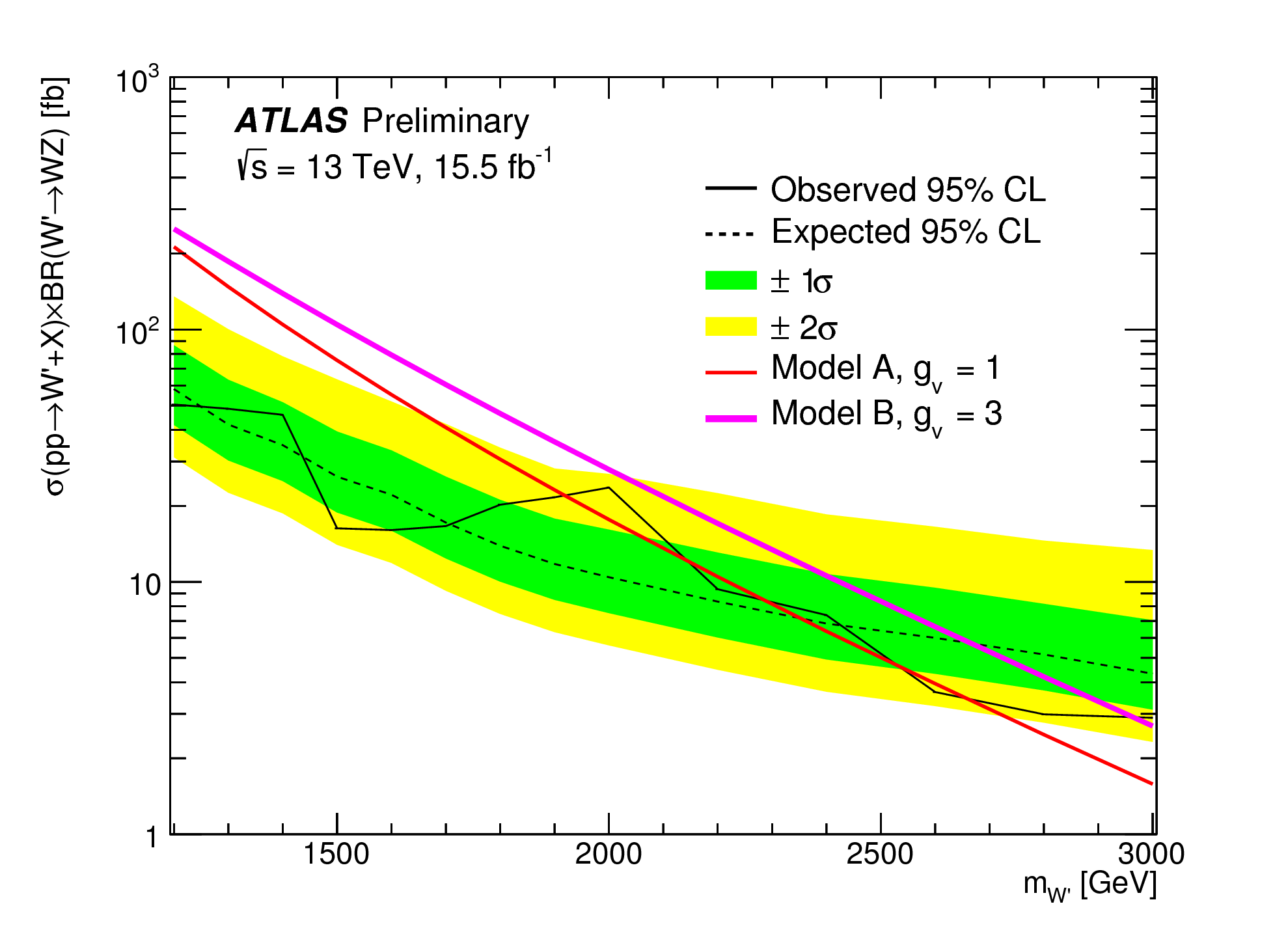}
\includegraphics[height=2.2in]{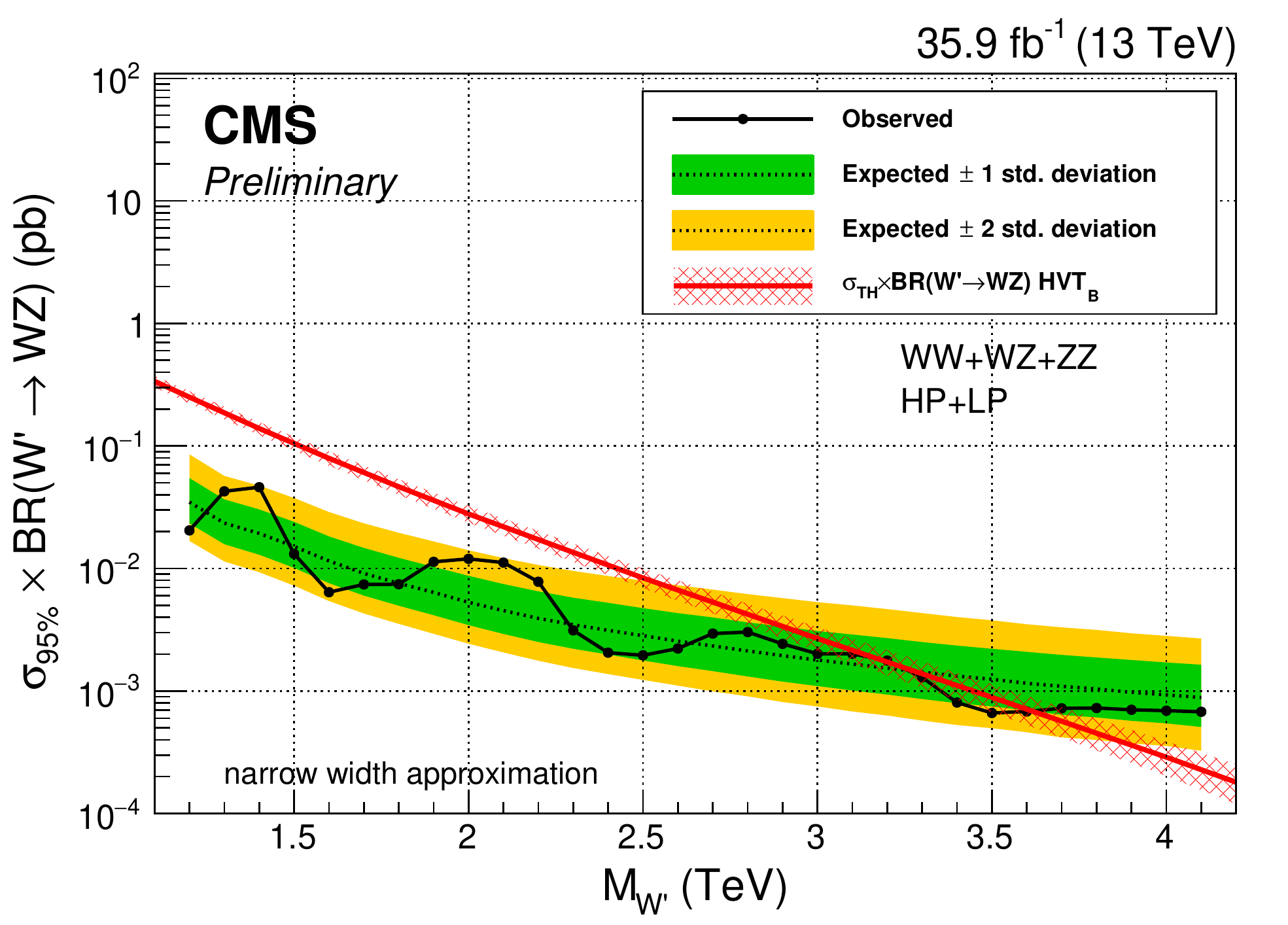}
\caption{ Observed (solid black) and expected (dashed) limits at
  95\% CL on the cross section times branching ratio for the W'$\rightarrow $ WZ search in ATLAS (left)
  \cite{ATLVV} and CMS
  (right) \cite{CMSVV}. The red and magenta curves show the predicted
  cross-sections as a function of resonance mass for the HVT model.}
\label{fig:VV}
\end{figure}

Searches for a WW/WZ/ZZ resonance in a fully hadronic final state have been
performed in both ATLAS \cite{ATLVV} and CMS \cite{CMSVV} using techniques and event
categorizations very similar to the ones used in VH searches in Sec. \ref{sec:VH}.
The multijet background is modelled with a parametric function and
validated in the sidebands of the V-candidate mass. No significant
excess is found and upper limits are set on production cross section times the
branching ratio to VV final states in $1.2<m_{JJ}<3$ TeV
($1.2<m_{JJ}<4.2$ TeV) range by ATLAS (CMS) (Fig. \ref{fig:VV}).

\section{ZZ $\rightarrow$ 2$\ell$2$\nu$}
\begin{figure}[htb]
%\begin{wrapfigure}[20]{r}{0.5\textwidth}
\centering
\includegraphics[width=0.42\textwidth]{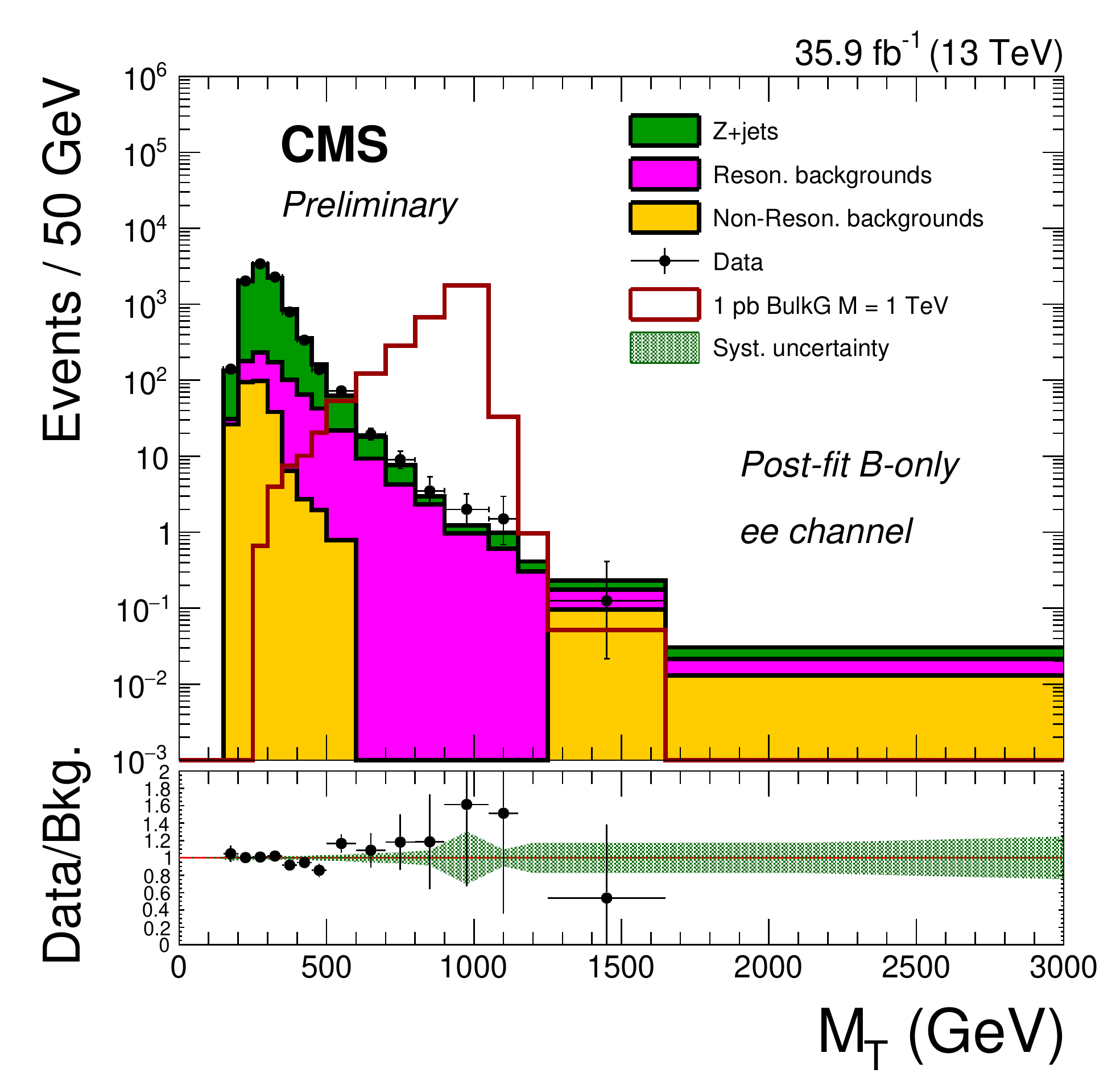}
\includegraphics[width=0.42\textwidth]{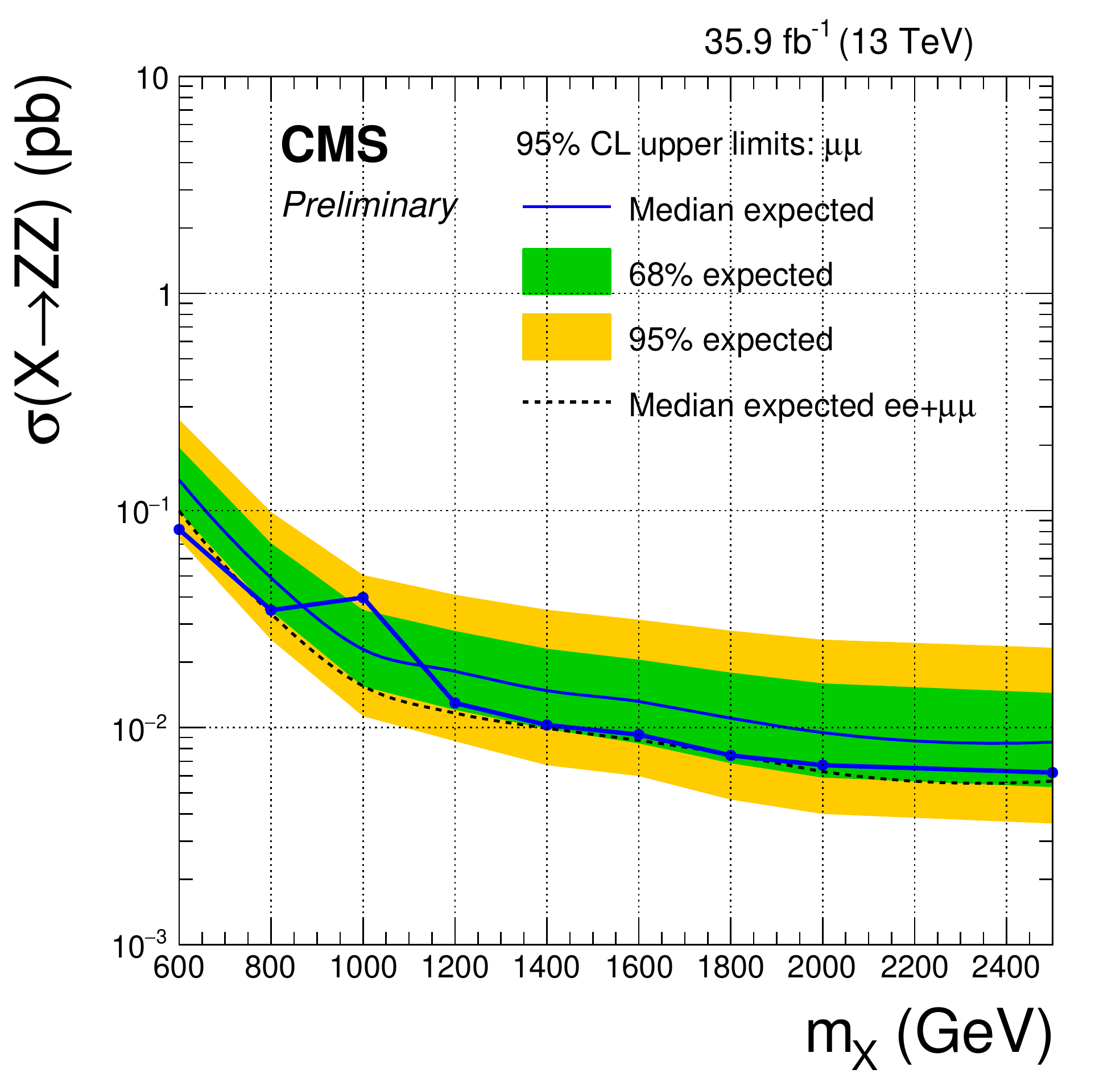}
\caption{ Left: The $M_T$ distribution for the electron channel in the
  search region comparing
  data and data-driven background modelling. Right: Observed (thick blue) and expected (thin blue) limits at
  95\% CL on the cross section times branching ratio for the $X\rightarrow $ ZZ search in CMS
 \cite{Zllvv}. }
\label{fig:Zllvv}
\end{figure}

This CMS search \cite{Zllvv} profits from a large branching ratio
of ZZ to 2$\ell$2$\nu$ (where $\ell =  e$ or $\mu$) final state and controllable backgrounds.
The main discriminating variable is the transverse mass ($M_T$) of the
two leptons and missing transverse momentum (denoted as $E_T^{\textrm{miss}}$), where a Jacobian edge from a potential resonance would be
expected in the distribution (Fig. \ref{fig:Zllvv}, left). 
The dominant background process is Z+jets production in which
$E_T^{\textrm{\small{miss}}}$ comes from mismeasurements of jet or lepton
momenta. It is estimated from a $\gamma$+jets data sample reweighted to
reproduce the kinematics of Z+jets events. The limits on
$\sigma(X\rightarrow ZZ)$ are presented
in Fig. \ref{fig:Zllvv} (right).

%See Figure \ref{fig:figure1} and Table \ref{tab:table1}. 
%%%%%%%%%%%%%%%%%%%%%%%%%%%%%%%%%%%%%%%%%%%%%%%%%%%%%%%%%%%%%%%%%%%%%%%%%
%%
%%   use this format to include an .eps figure into your paper
%\end{wrapfigure}
%%%%%%%%%%%%%%%%%%%%%%%%%%%%%%%%%%%%%%%%%%%%%%%%%%%%%%%%%%%%%%%%%%%%%%%%%%%

\section{HH $\rightarrow$ 4b}
\begin{figure}[htb]
\centering
\includegraphics[height=2.2in]{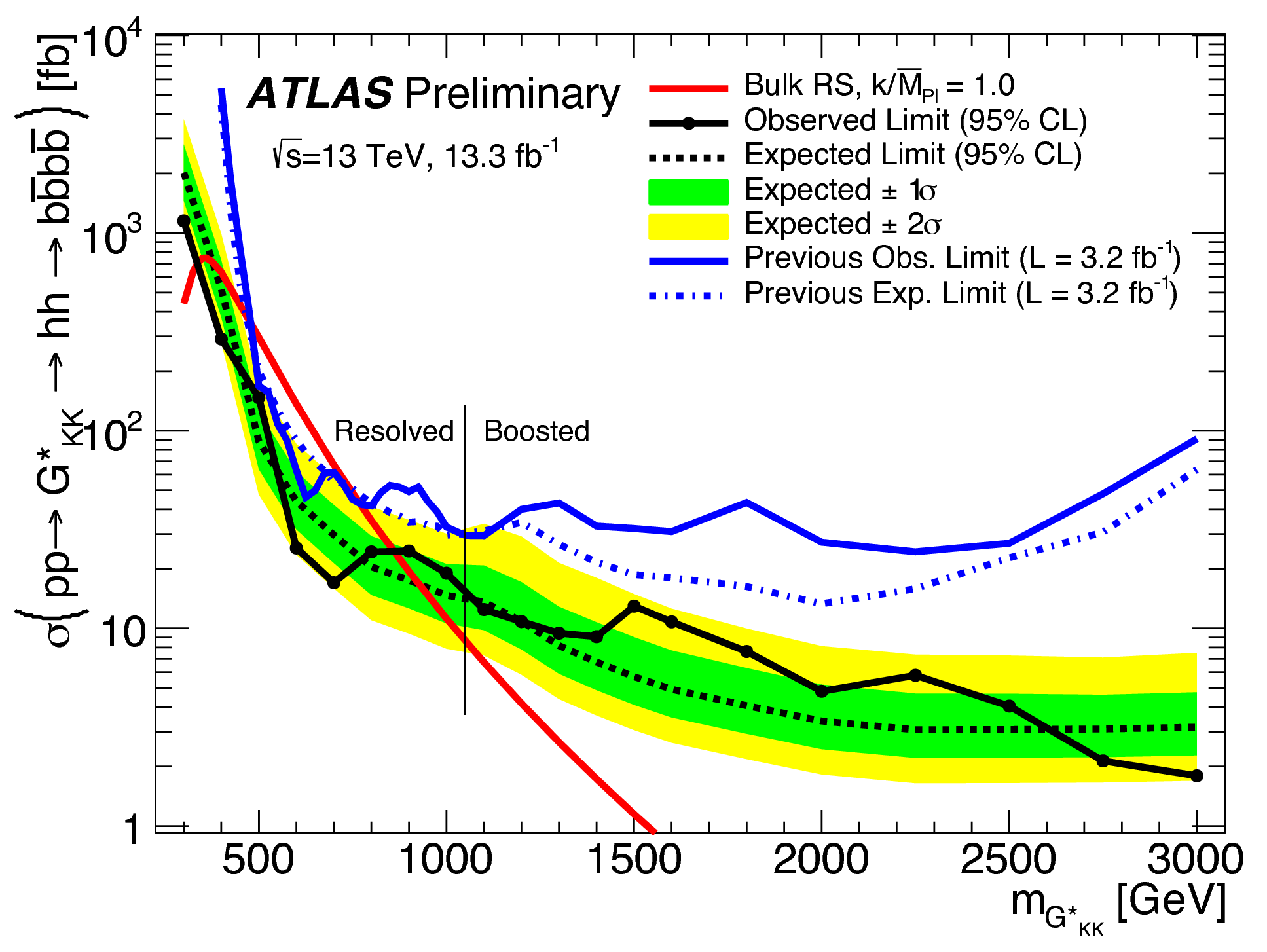}
\includegraphics[height=2.3in]{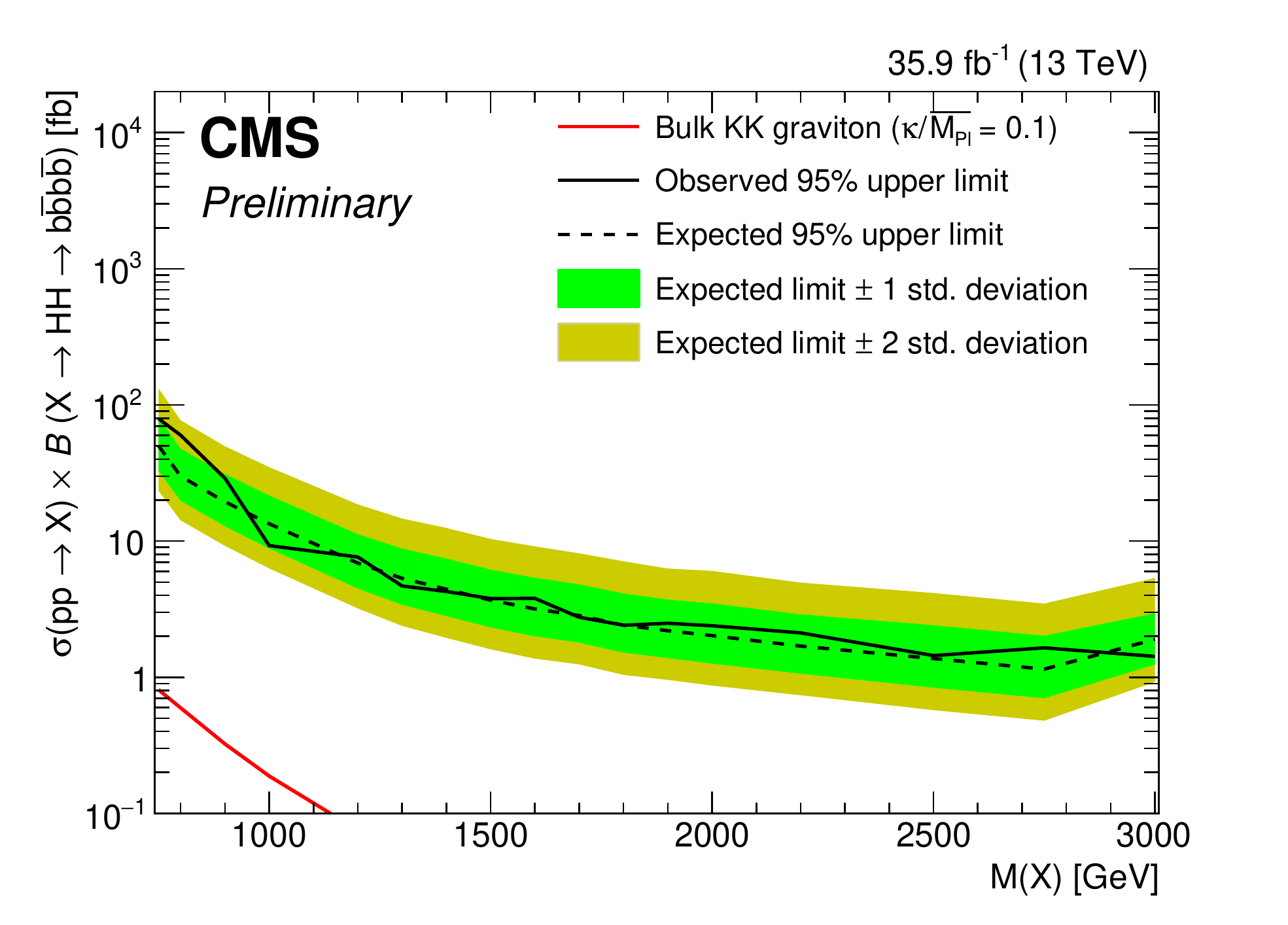}
\caption{ Observed (solid black) and expected (dashed) limits at
  95\% CL on the cross section times branching ratio for the X$\rightarrow $ HH search in ATLAS (left)
  \cite{ATLHH4b} and CMS
  (right) \cite{CMSHH4b}. The red curves show the predicted
  cross-sections as a function of resonance mass for the bulk
  Randall-Sundrum model.}
\label{fig:HH}
\end{figure}

The ATLAS search for a di-Higgs resonance  \cite{ATLHH4b} considers
both the resolved regime, where
four b-tagged  jets with R=0.4 can be reconstructed, and the boosted regime, where two fat
H-tagged jets are reconstructed. The resolved analysis provides
better sensitivity in the low-mass region, where the reconstructed
Higgs mass ($m_{HH}$) is below 1 TeV, while the boosted
analysis performs better at $m_{HH}>1$ TeV. The dominant multijet
background is modelled using an independent data sample -- from a region with exactly
two b-tagged jets (instead of at least four in the search region) in the resolved case and
from a region where none of the track jets associated to a H-candidate
were b-tagged in the boosted case.
The upper limits on the
production cross section times the branching ratio are set in the
context of the bulk Kaluza-Klein graviton ($G_{KK}^*$) and shown in
Fig. \ref{fig:HH} (left).

The CMS analysis \cite{CMSHH4b} requires two H-tagged fat jets with $M_{JJ}>$750
GeV. Events are separated into two categories depending on the
double-b-tag discriminator: events with tighter b-tagging criteria
having higher purity, but lower signal efficiency, and events with
looser b-tagging criteria and correspondingly lower purity but higher signal efficiency. The dominant
multijet background is estimated using the data from the region with
the inverted value of the b-tag discriminator. No significant excess is
observed and limits are set on the product of the production cross section
and the branching ratio in the $750<M_{JJ}<3000$ GeV range
(Fig. \ref{fig:HH}, right).

\section{Conclusions}

Extensive searches for high-mass diboson resonances have been
performed both in ATLAS and CMS.  So far no significant discrepancy
has been found between the observed data and the SM expectation. The
largest deviation from the background-only hypothesis has been seen in the ATLAS VH$\rightarrow$qqbb
search at $m_{VH} \sim 3$ TeV and has the global significance
of $\sim 2.2$ $\sigma$. The excess has not been observed
in the corresponding CMS analysis.

%%  if necessary
%\Acknowledgements

\end{document}